\documentclass[usenatbib]{mn2e}
\pdfoutput=1 
\usepackage{amsmath}
\usepackage{amssymb}
\usepackage{graphicx}
\usepackage{txfonts}
\usepackage{natbib}
\usepackage{wrapfig}
\usepackage{longtable}
\usepackage[usenames]{color}

\def\planck{{\it Planck}}
\def\Pcos{P16}
\def\Pccos{P20}
\def\Pclust{P29}
\def\WtG{{\it Weighing the Giants}}
\def\spose#1{\hbox to 0pt{#1\hss}}
\def\approxlt{\mathrel{\spose{\lower 3pt\hbox{$\sim$}}
        \raise 2.0pt\hbox{$<$}}}
\def\approxgt{\mathrel{\spose{\lower 3pt\hbox{$\sim$}}
        \raise 2.0pt\hbox{$>$}}}
\def\approxpropto{\mathrel{\spose{\lower 3pt\hbox{$\sim$}}
        \raise 2.0pt\hbox{$\propto$}}}

\bibpunct{(}{)}{;}{a}{}{,}

\title
[Planck cluster masses]
{Robust weak-lensing mass calibration of Planck galaxy clusters}
\author[Anja von der Linden et al.]
{\parbox[t]{\textwidth}{\vspace{-0.7cm}
\begin{flushleft}
Anja von der Linden$^{1,2,3}$
\thanks{E-mail:anja@slac.stanford.edu}, 
Adam Mantz$^{4,5}$,
Steven W. Allen$^{2,3,6}$,
Douglas E. Applegate$^{7}$,
Patrick L. Kelly$^{8}$,
R. Glenn Morris$^{2,6}$,
Adam Wright$^{2,3,6}$,
Mark T. Allen$^{2,3}$,
Patricia R. Burchat$^{2,3}$,
David L. Burke$^{2,6}$,
David Donovan$^{9}$,
Harald Ebeling$^{9}$
\end{flushleft}
}
\\
        \vspace*{2pt}
\\
$^{1}$Dark Cosmology Centre,  
Niels Bohr Institute, University of Copenhagen
Juliane Maries Vej 30,
2100 Copenhagen {\O},
Denmark \\
$^{2}$Kavli Institute for Particle Astrophysics and Cosmology,
Stanford University,
452 Lomita Mall,
Stanford, CA  94305-4085, USA\\
$^{3}$Department of Physics,
Stanford University,
382 Via Pueblo Mall, 
Stanford, CA  94305-4060, USA\\
$^{4}$Kavli Institute for Cosmological Physics,
University of Chicago,
5640 South Ellis Avenue,
Chicago, IL 60637-1433, USA\\
$^{5}$Department of Astronomy and Astrophysics, 
University of Chicago,
5640 South Ellis Avenue,
Chicago, IL 60637-1433, USA\\
$^{6}$SLAC National Accelerator Laboratory, 
2575 Sand Hill Road, 
Menlo Park, CA 94025, USA\\
$^{7}$Argelander-Institut f\"ur Astronomie,
Universit\"at Bonn, 
Auf dem H\"ugel 71, 
53121 Bonn, Germany\\
$^{8}$Department of Astronomy, 
University of California, 
B-20 Hearst Field Annex \# 3411,
Berkeley, CA 94720-3411, 
USA\\
$^{9}$Institute for Astronomy, 
2680 Woodlawn Drive, 
Honolulu, HI 96822, USA\\
}

\begin{document}

\date{}

\pagerange{\pageref{firstpage}--\pageref{lastpage}} \pubyear{2014}

\maketitle

\label{firstpage}
\vspace{-1cm}
\begin{abstract}
\noindent
In light of the tension in cosmological constraints reported by the
\planck\ team between their SZ-selected cluster counts and Cosmic
Microwave Background (CMB) temperature anisotropies, we compare the
\planck\ cluster mass estimates with robust, weak-lensing mass
measurements from the {\it Weighing the Giants} (WtG) project.  For
the 22 clusters in common between the \planck\ cosmology sample and
WtG, we find an overall mass ratio of $\left< M_{Planck}/M_{\rm WtG}
\right> = 0.688 \pm 0.072$.  Extending the sample to clusters not used
in the \planck\ cosmology analysis yields a consistent value of
$\left< M_{Planck}/M_{\rm WtG} \right> = 0.698 \pm 0.062$ from 38
clusters in common.  Identifying the weak-lensing masses as proxies
for the true cluster mass (on average), these ratios are
$\sim\!\!1.6\sigma$ lower than the default bias factor of 0.8 assumed in
the \planck\ cluster analysis.  Adopting the WtG weak-lensing-based
mass calibration would substantially reduce the tension found between
the \planck\ cluster count cosmology results and those from CMB
temperature anisotropies, thereby dispensing of the need for
  ``new physics'' such as uncomfortably large neutrino masses (in the
  context of the measured \planck\ temperature anisotropies and other
  data).  We also find modest evidence (at 95 per cent confidence)
for a mass dependence of the calibration ratio and discuss its
potential origin in light of systematic uncertainties in the
temperature calibration of the X-ray measurements used to calibrate
the \planck\ cluster masses.  Our results exemplify the critical role
that robust absolute mass calibration plays in cluster cosmology, and
the invaluable role of accurate weak-lensing mass measurements in this
regard.
\end{abstract}


\section{Introduction}
\label{sect:intro}

The \planck\ satellite has recently provided new, precise cosmological
constraints based on measurements of Cosmic Microwave Background (CMB)
temperature anisotropies \citep[][hereafter \Pcos]{Planck13_16},
confirming that a spatially flat $\Lambda$CDM model provides an
excellent description of the observable Universe. However, an
uncomfortable result of the \planck\ team's analysis is that for a
spatially flat $\Lambda$CDM model, their cosmological constraints
derived from the number density of galaxy clusters detected with
\planck\ through the Sunyaev-Zel'dovich (SZ) effect are in tension
with the constraints from the CMB temperature power spectrum: in
particular, the \planck\ cluster analysis \citep[][hereafter
  \Pccos]{Planck13_20} prefers a significantly lower value of the
amplitude of the matter power spectrum at late times, $\sigma_8$,
compared to the value extrapolated from CMB temperature anisotropies.

The \planck\ team suggest two possible explanations for this tension.
One is that the calibration of their cluster mass estimates, which are
based on hydrostatic mass measurements derived from {\it XMM-Newton}
X-ray observations, is biased low. An important aspect of that
analysis is the assumed value of the mass bias, $\left<
M_{Planck}/M_{\rm true} \right> = (1-b)$.  Expecting some level of
non-thermal pressure support to be present in even relatively relaxed
clusters, they present two sets of results, one which sets
$(1-b)\equiv0.8$, and one which marginalizes over the range $0.7 <
(1-b) < 1.0$.

Reconciling the observed tension between \planck\ SZ cluster counts
and the CMB power spectrum would, however, require an even lower
ratio.  Alternatively, keeping the \planck\ mass calibration fixed at
$(1-b) = 0.8$, the \planck\ team argue that their CMB and SZ-cluster
data could be explained by a species-summed neutrino mass of $\Sigma
m_{\nu} = (0.58 \pm 0.20)$~eV ($2.8\sigma$ significance departure from
zero), a result that is in tension with the 95 per cent confidence
upper limit of 0.23~eV derived from the \planck\ CMB analysis (\Pcos)
in combination with WMAP low-multipole polarization \citep{blw13},
high-multipole temperature anisotropy from SPT and ACT
\citep{rsz12,dln14}, and the 6df, SDSS and BOSS BAO surveys
\citep{bbc11,pxe12,aab12}, and earlier studies using WMAP CMB data and
independent X-ray and optical cluster measurements
\citep{mar10c,rvj10}.  Marginalizing over $0.7 < (1-b) < 1.0$, the
evidence for massive neutrinos is weakened but still present, $\Sigma
m_{\nu} = (0.40 \pm 0.21)$~eV ($1.9\sigma$).

In this letter, we investigate the reliability of the \planck\ cluster
mass measurements by comparing them with robust, independently derived
weak-lensing mass measurements from the \WtG\ (WtG) project
\citep{laa14, kla14,alk14}.  As has been discussed in the literature,
systematic uncertainty in the absolute calibration of cluster masses
is currently the most significant challenge facing cluster cosmology
\citep[WtG; see also][]{vkb09,mar10,rwr10,sta11,bhd13,aem11,lsst12}.
Weak lensing provides our most promising method to calibrate the
absolute masses of clusters since it measures the total mass directly,
without relying on baryonic tracers, and is expected, from
simulations, to be accurate (i.e. exhibit minimal bias in the mean).
The WtG project targeted a subset of 51 clusters in catalogs based on
the ROSAT All-Sky Survey \citep[RASS,][]{tru93}. The overlap between
those X-ray detected clusters and the \planck\ cluster catalog
\citep[see ][hereafter \Pclust]{Planck13_29} is substantial: 38 WtG
clusters are also in the \planck\ cluster sample, 22 of which were
used in the \planck\ cluster count cosmology analysis. The WtG masses
hence provide an excellent external, independent dataset to assess the
calibration of the \planck\ cluster masses.

The masses and mass ratios quoted in this paper assume a flat
$\Lambda$CDM cosmological model with $\Omega_{\rm m} = 0.3$ and $H_0 =
100 \,h\, \mbox{km/s/Mpc}$, where $h=0.7$.

\section{Data}
\label{sect:data}

We compare \planck\ and WtG measurements of $M_{500}$, i.e. the
integrated mass inside the associated radius $r_{500}$, within which
the average density is 500 times the critical density at the cluster
redshift. We note that because the mass is defined as enclosing a
given mean density (rather than a given physical radius), the impact
of any systematic calibration uncertainty in these measurements is
boosted in the comparison: if the masses are systematically
underestimated, the $r_{500}$ values are also underestimated, leading
to even lower cluster mass estimates.  Nevertheless, direct comparison
of the $M_{500}$ mass estimates is sensible, since these are the
relevant quantities for the \planck\ cluster cosmology analysis.
The \planck\ $M_{500}$ estimates are listed in Table C.1 of \Pclust.

\subsection{Unbiased weak-lensing mass estimates from {\it Weighing the Giants}} 

A key assumption in the present analysis is that the WTG weak-lensing
cluster mass estimates are unbiased {\it on average}.  In
\citet{laa14} and \citet{alk14}, we discuss potential residual sources
of bias and quantify the uncertainties that they introduce. We provide
a short summary here, and refer the reader to the WtG papers for more
detailed explanations.

Weak lensing cluster mass estimates require three components:
estimates of the distortion of each galaxy image due to the cluster
(``shear''), estimates of the redshifts of the galaxies used in the
weak lensing analysis (since the shear induced on a background galaxy
depends on the ratio of angular diameter distances between the
observer, cluster, and source), and an assumption regarding the mass
distribution of the cluster.  

We calibrate our shear estimates using the STEP 1 and 2 simulations
\citep{hvb06,mhb07}, which are ideally designed to calibrate shear
measurements in Subaru images.  Because the STEP simulations focus on
relatively small shear values and non-crowded fields -- assumptions
which cannot be extended to cluster cores -- we restrict the analysis
to radii $>0.75$\,Mpc for the mass measurements.

In terms of the redshifts / redshift distribution of the galaxies in
the cluster fields, we estimate cluster masses with two different
methods in \citet{alk14}.  The standard ``color-cut'' method
\citep{hoe07,otu10} which can be applied to observations in just two
bands (or even a single band), implicitly places all galaxies at a
single ``effective'' redshift, which is estimated by applying the same
cuts in magnitude, colors, etc. to both the cluster field and to a
reference deep field, with dilution due to cluster members not on the
red sequence corrected using simple number density profiles. For the
highest precision measurements, and especially at increasing cluster
redshift, both the deep-field assumption and the cluster member
correction are expected to break down due to cosmic variance,
observational selection effects, etc.  To overcome this limitation and
provide improved robustness, we develop a maximum-likelihood method to
measure weak-lensing cluster masses based on full photometric redshift
probability distributions for individual galaxies.  With extensive
simulations we show that, when restricted to an optimal redshift
range, this method yields nearly unbiased cluster mass measurements.
Since not all WtG clusters have been imaged in sufficient bands to
provide robust photometric redshift estimates, we use the subset of 27
clusters with full 5-filter photometry and masses determined with the
improved method to calibrate traditional ``color-cut'' masses for the
full sample of 51 clusters. In this way we obtain the largest overlap
possible with the \planck\ analysis.

Finally, since lensing is sensitive to all matter along the line of
sight and thus best at measuring projected (cylindrical) masses, some
assumption about the cluster mass distribution needs to be made in order
to infer a spherical overdensity mass such as $M_{500}$ from the
weak-lensing measurements.  For our analysis, we fit a spherically
symmetric NFW model to the observed shear profiles around cluster
centroids determined from {\it Chandra} imaging in order to determine
$M_{500}$.  Since clusters are generally triaxial, have substructure,
and other structures projected along the line of sight, we expect
significant scatter in the relation between the ``true'' mass, and the
mass inferred from lensing.  From simulations, this intrinsic scatter
is expected to be of the order of 20\%; in ground-based observations
such as those considered here, the shot noise in the shear estimates
due to the intrinsic ellipticity of galaxies adds a similar
statistical scatter, yielding a total expected scatter of $\sim 30$\%
\citep{bek11}.  Therefore, relatively large samples of clusters -
selected in a way that does not depend on their lensing properties -
are required to determine the mass calibration to high precision.  The
key question then becomes whether the {\it average} weak-lensing mass
is unbiased. Fortunately, simulations indicate that for the most
massive clusters, the average mass from an NFW fit is unbiased to
within a few percent if the radial fit range is restricted to be
within the virial radius \citep{bek11,bmk11,ogh11,gmm14}.  Given that
the clusters considered here are among the most massive known, we fit
our shear profiles over the range of 0.75--3\,Mpc.  The inner cut-off
ensures that the quoted mass is largely insensitive to the choice of
the concentration of the NFW profile: we assume $c=4$, appropriate for
the most massive clusters, and have verified that even substantial
shifts in the assumed concentration cause only slight shifts in the
resulting masses.

In \citet{alk14}, we quantify the systematic uncertainty associated
with each of these components and show that the WTG data for 51
clusters determine the ensemble mean mass to a systematic precision of
7~per~cent (8~per~cent when statistical uncertainties are included).

Note that the X-ray-selection of the WtG clusters ensures that the
comparison sample is fair (at least to the extent required here):
selection by X-ray luminosity is largely insensitive to triaxiality
and orientation along the line of sight \citep[e.g.][]{aem11}.
Instead, for X-ray selection, the dominant source of scatter is the
existence/absence of a cool core.  The WtG sample is thus unbiased in
the sense that for a given 'true' cluster mass, it is equally likely
to have selected a cluster that scatters 'up' or 'down' in
weak-lensing mass.

\begin{figure}
\includegraphics[width=\hsize]{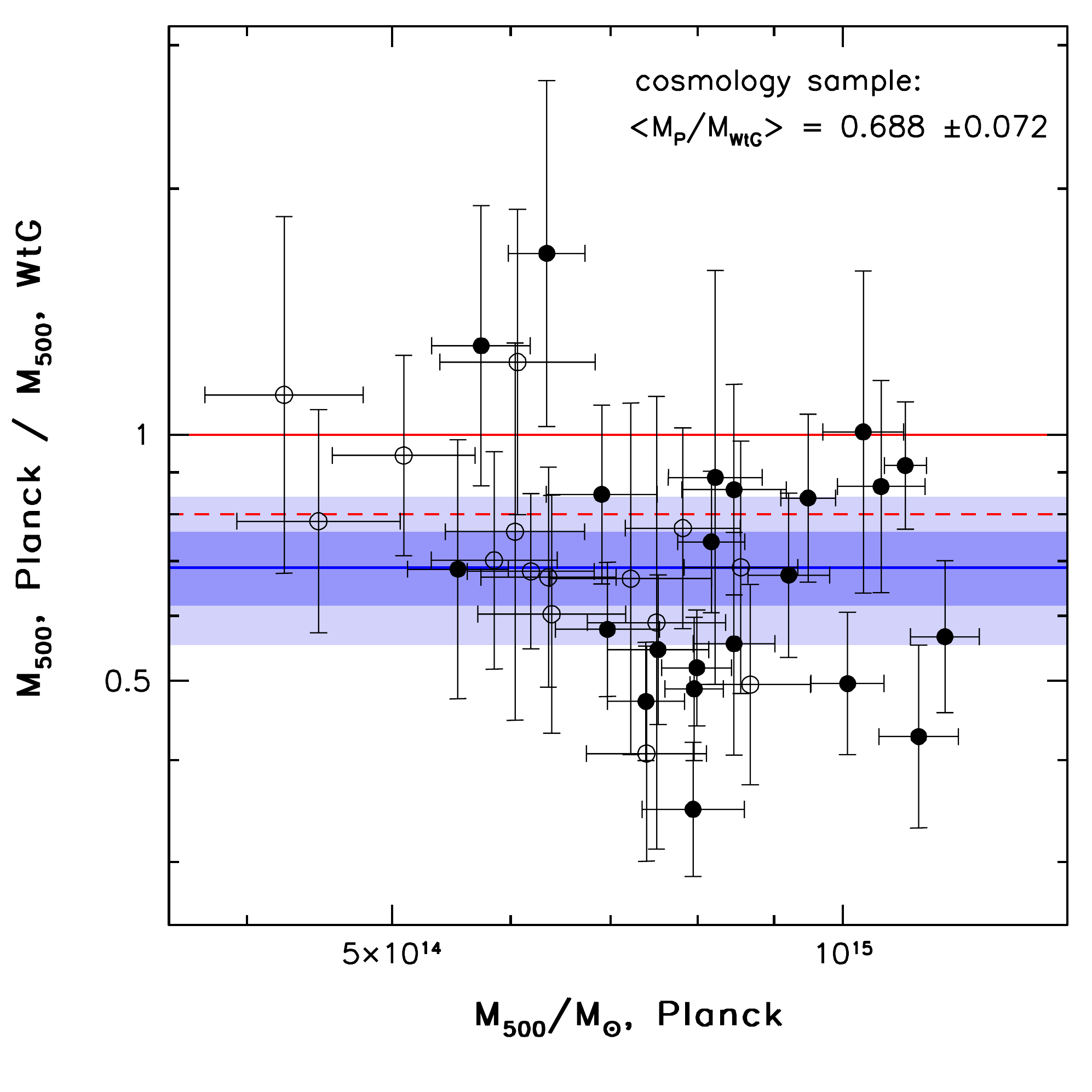}
\caption{The ratio of cluster masses measured by \planck\ and by WtG,
  for the clusters common to both projects.  Solid symbols denote
  clusters which were included in the \planck\ cluster cosmology
  analysis (22 clusters) and open symbols additional clusters in the
  \planck\ cluster catalog (16 clusters).  The red, solid line
  indicates a ratio of unity (no bias). The dashed red line indicates
  $(1-b)=0.8$, the default value assumed throughout most of \Pcos.
  The blue line and shaded regions show our best-fit mass ratio along
  with the 1- and 2-$\sigma$ confidence intervals.  Since the
  weak-lensing masses are expected to be unbiased {\it on average},
  the ratio of \planck\ masses to weak-lensing masses is a measure of
  the bias $(1-b)=M_{Planck}/M_{\rm true}$ of the \planck\ cluster
  masses as used in \Pccos. }
\label{fig:ratio}
\end{figure}

\begin{figure}
\includegraphics[width=\hsize]{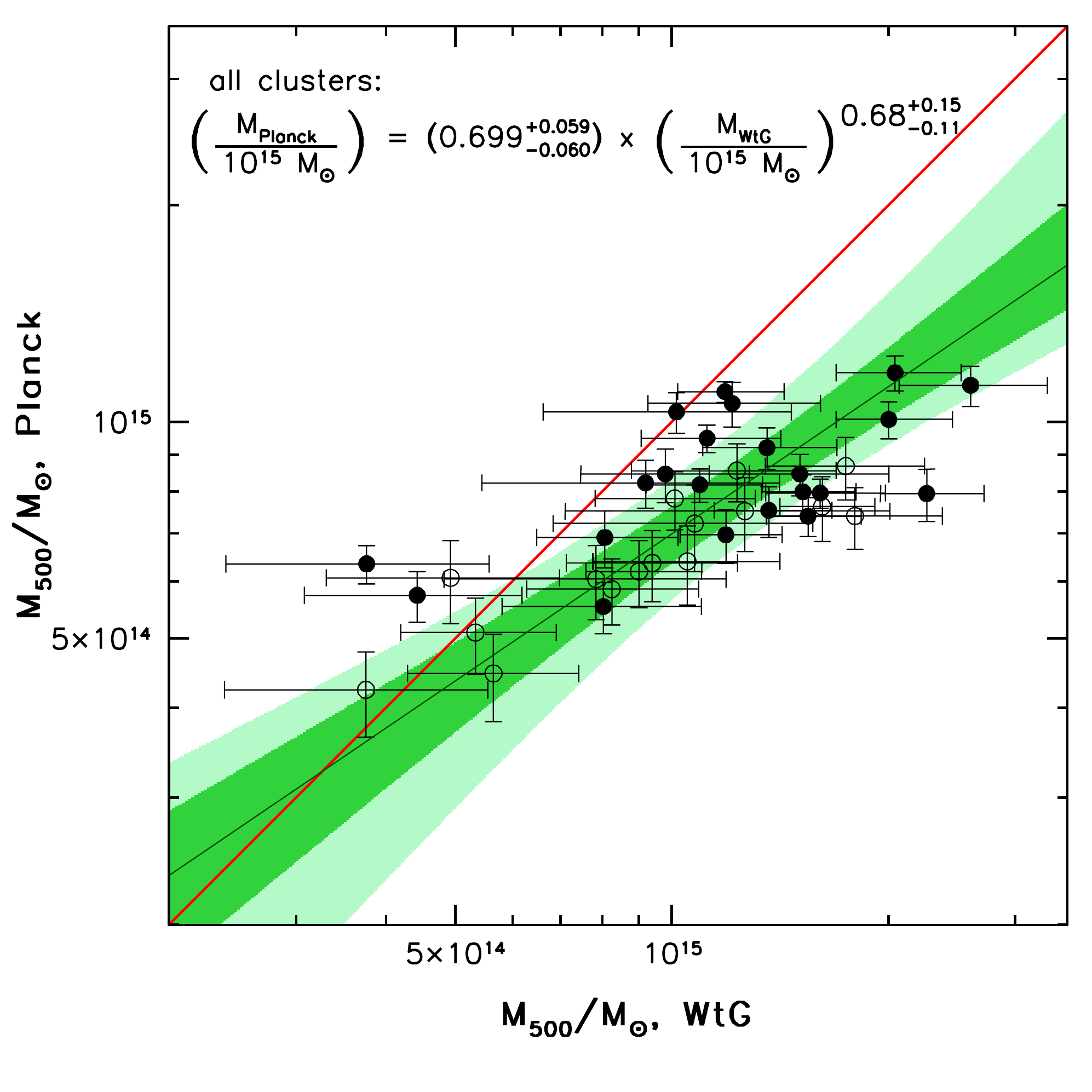}
\caption{The direct comparison between $M_{500}$ cluster masses
  measured by \planck\ and by WtG.  The symbols are the same as in
  Fig.~\ref{fig:ratio}. The green line and shaded regions show the
  best-fit linear relation between the logarithmic masses and its 1-
  and 2-$\sigma$ confidence intervals (the fit was performed with
  $\log(M_{\rm WtG})$ as function of $\log(M_{Planck})$). }
\label{fig:masses}
\end{figure}

\section{Results}
\label{sect:results}

\subsection{Average mass calibration}

In Fig.~\ref{fig:ratio}, we show the ratio $M_{Planck}/M_{\rm WtG}$ as
a function of $M_{Planck}$.  The \planck\ mass estimates are
considerably lower than the weak-lensing based WtG masses, especially
at large \planck\ masses.  Using bootstrap realizations of the
unweighted mean\footnote{We verified that this procedure returns
  unbiased estimators of the mean and standard deviation of a
  log-normal distribution, even in the presence of intrinsic scatter
  and if the measurement uncertainties correlate with the measured
  values (as can be seen in Fig.~\ref{fig:ratio}, less massive
  clusters have larger error bars and higher $M_{Planck}/M_{\rm
    WtG}$).}, we find the average mass ratio for the 22 clusters used
in the \planck\ cluster cosmology analysis to be:
$$ 
\beta_{\rm cos} = \left< \frac{M_{Planck}}{M_{\rm WtG}} \right> = 0.688^{+0.056}_{-0.050} \,{\rm (stat)} \pm
0.049\, {\rm (syst)} \quad .
$$ 
The systematic uncertainty quoted here
expresses the systematic uncertainty on the weak-lensing masses,
i.e. it includes all entries in Table~4 of \citet{alk14} with the
exception of the scatter due to triaxiality, which is accounted for
here in the statistical uncertainty. Extending the sample to all 38
clusters yields a consistent result:
$$ 
\beta_{\rm all} = 0.698^{+0.039}_{-0.037} \,{\rm (stat)} \pm 0.049\, {\rm (syst)} \quad .
$$ 

The weak-lensing masses are expected to yield the {\it true} cluster
mass on average, and thereby enable a robust calibration of other mass
proxies \citep[see discussion in][]{laa14,alk14}.  Therefore, by
identifying $\beta = (1-b)$, these results suggest that the mass
calibration adopted by the Planck team, $(1-b)\equiv0.8$,
underestimates the true cluster masses by between 5 and 25 per cent on
average.  Similarly, their alternative analysis, which
  marginalizes over $0.7 < (1-b) < 1.0$ is significantly skewed from
  the range preferred by our weak-lensing calibration.

\subsection{Evidence for a mass-dependent calibration problem}

By eye, Fig.~\ref{fig:ratio} suggests that the ratio between the WtG
weak-lensing and \planck\ mass estimates depends on the cluster mass:
at masses $\lesssim 6\times10^{14} M_{\odot}$, the mass estimates
roughly agree, whereas the discrepancy appears significant for more
massive clusters.  To quantify the evidence for such a mass-dependent
bias, we use the Bayesian linear regression method developed by
\citet{kel07} to fit $\log(M_{\rm WtG})$ as function of
$\log(M_{Planck})$. Fitting the masses directly avoids the correlated
errors in the mass ratios one would have to account for if fitting the
data as shown in Fig.~\ref{fig:ratio}.  While we show $M_{Planck}$ as
function of $M_{\rm WtG}$ in Fig.~\ref{fig:masses} to reflect that the
weak-lensing masses are our proxy for true cluster masses, we assign
the \planck\ mass estimates to be the independent variable of the fit
to reduce the effects of Malmquist bias: $M_{Planck}$ scales with the
survey observable, and by choosing it as the independent variable, we
provide a mass estimate for each data point which is to first order
independent of other selection effects (as X-ray selection to first
order does not correlate with SZ selection biases, and the lensing
data are a subsample of an X-ray selected catalog).  The \citet{kel07}
method accounts for measurement errors in both variables, as well as
for intrinsic scatter in the dependent variable. Rephrasing the
results as a power-law, the best-fit relation for the 22 clusters in
the cosmology sample is:
$$
\left( \frac{M_{Planck}}{10^{15} M_{\odot}} \right) = \left( 0.697^{+0.077}_{-0.095} \right) \times \left( \frac{M_{\rm WtG}}{10^{15} M_{\odot}} \right)^{0.76^{+0.39}_{-0.20}} \quad ,
$$ 
where the systematic uncertainty on the weak-lensing mass calibration is
accounted for in the uncertainty on the coefficient. 
In 24 per cent of the
Monte Carlo samples, the slope (of $\log(M_{Planck})$ vs. $\log(M_{\rm
  WtG})$) is unity or larger; i.e. the evidence for a mass-dependent
bias is at the $\sim1\sigma$ level for these 22 clusters.

To further test for a mass-dependent bias, it is instructive to
include the additional 16 clusters in common between \planck\ and WtG
that are not used in the \planck\ cluster cosmology analysis, as these
extend the mass range probed.  For all 38 clusters, we find a
consistent and more precise result:
$$
\left( \frac{M_{Planck}}{10^{15} M_{\odot}} \right) = \left( 0.699^{+0.059}_{-0.060} \right) \times \left( \frac{M_{\rm WtG}}{10^{15} M_{\odot}} \right)^{0.68^{+0.15}_{-0.11}} \; .
\label{eq:mp_v_ml}
$$
In 4.9 per cent of the Monte Carlo samples, the slope is unity or larger;
i.e. the confidence level for a slope less than unity is 95 per cent\footnote{We
note that when using bootstrap realizations of an unweighted simple
linear regression as a more agnostic fit statistic, we recover the
same slope, but with smaller uncertainties ($0.69^{+0.11}_{-0.08}$).
In only 1.7 per cent of the bootstrap samples is the slope larger than
unity. However, we choose to quote the more conservative error
estimates returned by the \citet{kel07} method.}.

\section{Discussion}

\subsection{Implications for cosmological constraints}

Comparing \planck\ cluster mass estimates (calibrated with hydrostatic
mass estimates from {\it XMM-Newton} data) with weak-lensing-based
mass measurements from the \WtG\ project, we have measured the bias of
the \planck\ masses to be $\beta_{\rm cos} = \left< M_{Planck}/M_{\rm
  true} \right > = (1-b) = 0.688 \pm 0.072$ for 22 clusters in the
\planck\ cosmology sample ($\beta_{\rm all} = 0.698 \pm 0.062$ for all
38 clusters in common between the two studies).  This result assumes
that the WtG mass measurements are unbiased to the level of accuracy
discussed in \citet{alk14}. Our result suggests that the default mass
calibration adopted by the Planck team, $(1-b)=0.8$, underestimates
the true masses of Planck clusters by 5--25 per cent on average.  More
than half of the probability distribution we find for $\beta = (1-b)$
is outside of the range of 0.7--1.0 that the \planck\ team marginalize
over in their most conservative analysis.

A significantly lower value for the
mass bias would reduce the tension between their SZ cluster count and
CMB analyses \citep[\Pccos, also see][]{rrb14,rer14,rbe14}.  Adopting
the WtG mass calibration, which includes a sizable range of lower
values for the mass bias would therefore reduce the tension and bring
the constraints into significantly better agreement.
In particular, it would remove the need to evoke unusually large
  neutrino masses in the context of a purely cosmological explanation
  of the claimed tension.

\subsection{On the origin of the mass bias}

The mass estimates used by the \planck\ team are based on hydrostatic
modeling of {\it XMM-Newton} X-ray data for 20 relatively relaxed
clusters \citep{app10}. The mass bias term, $1-b$, is primarily
intended to account for departures from hydrostatic equilibrium (the
`hydrostatic bias'). In practice, the bias term must account for the
total, summed systematic offset between the mass estimates and true
masses. Here the relevant terms include not just the hydrostatic bias,
but also the effects of, e.g., instrument calibration, non-thermal
pressure support, and temperature inhomogeneities.

Hydrostatic biases at the 10--20 per cent level at $r_{500}$ (in the
sense that X-ray measurements underestimate the true mass) are
expected to exist even in relatively relaxed, massive clusters
\citep[e.g.][]{nvk07,rmm12}.  This expectation is reflected in the
choice of $(1-b)=0.8$ (or $0.7 < (1-b) < 1.0$) by the
\planck\ team.  On the calibration side, cluster temperature (and
therefore mass, $M\approxpropto T^{3/2}$) estimates from {\it
  XMM-Newton} are typically lower than {\it Chandra}-based values for
massive clusters: whereas {\it XMM-Newton} and {\it Chandra} cluster
temperature measurements agree well for relatively low-mass systems
(with temperatures $kT<2$~keV), for massive clusters ($kT\geq5$~keV)
typical of systems found in the \planck\ SZ survey, {\it XMM}-based
temperature estimates tend to be about 20 per cent lower than {\it
  Chandra} values \citep{ggn13,mhb13,srl14}. This implies that an {\it
  XMM}-based $M_{500}$ mass estimate for a $\sim 6$ keV cluster will
be $\sim 30$ per cent lower than the corresponding {\it Chandra}
value.  The temperature-dependent discrepancy between the two cautions
that mass proxies which incorporate X-ray temperature measurements
require careful mass-/temperature-dependent calibration efforts.  In
this context, the trend in the ratio of \planck\ and WtG masses,
implying larger bias at higher masses, is interesting.  In
combination, the effects of hydrostatic masses biases and
temperature-dependent systematic offsets can easily reach 30 per cent,
and provide a natural explanation for both the value of $\beta = (1-b)
= M_{Planck}/M_{\rm WtG}$ we observe, and the mass-dependent slope in
the logarithmic ratio.

We deem it unlikely that the WtG weak-lensing measurements have a
significant mass-dependent bias. Simulations indicate that
weak-lensing mass estimates derived by fitting an NFW profile over the
radial range used in the WtG study (which does not exceed the cluster
virial radii nor extend too close to the cluster centers) should be
nearly unbiased for systems in the mass range probed here
\citep{ogh11,bek11,bmk11}. Any remaining biases affecting the WtG
masses should be at most of the order of a few per cent.  We note also
that by comparing Planck $Y_{500}$ estimates with independent
weak-lensing mass estimates for five clusters, \citet{gsk13} report a
mass dependence of the mass ratio in the same sense as we find.

Our results highlight the need for robust weak-lensing measurements to
complement X-ray and SZ data in determining mass estimates for galaxy
clusters. Whereas X-ray measurements can provide precise, {\it
  low-scatter} mass proxies for clusters, exhibiting minimal
($<$10--15 per cent) intrinsic scatter with respect to true mass
\citep[e.g.][]{ase04,ars08,kvn06}, weak-lensing data can provide an
accurate {\it absolute} mass calibration of these proxies. In this
way, through the combination of multi-wavelength data, the dominant
systematic uncertainty currently affecting cluster cosmology can be
largely circumvented.

\section{Summary and conclusion}

In light of the reported tension between the cosmological constraints
inferred from \planck\ SZ cluster counts and \planck\ CMB temperature
anisotropies, we have compared the \planck\ team's cluster mass
estimates with weak-lensing-based mass measurements determined by the
\WtG\ project.  For the 22 clusters in common between the
\planck\ cluster sample and WtG that are used in the \planck\ cluster
cosmology analysis, we find the average mass ratio to be $\left<
M_{Planck}/M_{\rm WtG} \right> = (1-b) = 0.688 \pm 0.072$.  For all 38
clusters in common between the two studies, we find $\left<
M_{Planck}/M_{\rm WtG} \right> = 0.698 \pm 0.062$.  These values are
$\sim1.6\sigma$ below the default bias adopted by \Pccos, $(1-b)=0.8$.
Also the range over which they marginalize for their more
  conservative analysis (0.7--1.0) is significantly skewed compared to
  the WtG mass calibration -- half of the WtG range is not included.
Adopting the WtG mass calibration would raise the value of $\sigma_8$
inferred from the \planck\ SZ cluster counts and alleviate the tension
with the Planck CMB results.

We find modest (95 per cent) evidence for a mass-dependence of the
mass calibration, with a best-fit power-law of $M_{Planck} \propto
{M_{\rm WtG}}^{0.68^{+0.15}_{-0.11}}$.  As a possible origin for such
a mass dependence we identify the temperature-dependent calibration
uncertainty of the X-ray hydrostatic measurements used to calibrate
the \planck\ cluster mass estimates.

Weak-lensing measurements provide an excellent complement to X-ray
and SZ data in enabling the robust calibration of cluster masses.  
In forthcoming work (Mantz et al., in prep.), we will present
cosmological constraints utilizing the full WtG sample in conjunction
with X-ray selected cluster catalogs and extensive Chandra follow-up
data.

\section*{Acknowledgements}

We thank the referee for helpful suggestions to clarify the
paper. Based in part on data collected at Subaru Telescope (University
of Tokyo) and obtained from the SMOKA, which is operated by the
Astronomy Data Center, National Astronomical Observatory of Japan.
Based on observations obtained with MegaPrime/MegaCam, a joint project
of CFHT and CEA/DAPNIA, at the Canada-France-Hawaii Telescope (CFHT)
which is operated by the National Research Council (NRC) of Canada,
the Institute National des Sciences de l'Univers of the Centre
National de la Recherche Scientifique of France, and the University of
Hawaii.  This research has made use of the SZ-Cluster Database
operated by the Integrated Data and Operation Center (IDOC) at the
Institut d'Astrophysique Spatiale (IAS) under contract with CNES and
CNRS.

This work was supported in part by the National Science Foundation
under grants AST-0838187 and AST-1140019, the U.S. Department of
Energy under contract number DE-AC02-76SF00515, the German Federal
Ministry of Economics and Technology (BMWi) under project 50 OR 1210,
and the National Aeronautics and Space Administration (NASA) through
Chandra Award Numbers GO8-9118X and TM1-12010X (issued by the Chandra
X-ray Observatory Center, which is operated by the Smithsonian
Astrophysical Observatory for and on behalf of NASA under contract
NAS8-03060) and program HST-AR-12654.01-A, provided by NASA through a
grant from the Space Telescope Science Institute (operated by the
Association of Universities for Research in Astronomy, Inc., under
NASA contract NAS 5-26555).  The Dark Cosmology Centre (DARK) is
funded by the Danish National Research Foundation.

\bibliography{../weaklensingpapers/refs.bib,refs.bib}

\label{lastpage}

\end{document}